 \documentclass[final,5p,times,twocolumn,authoryear]{elsarticle}

\usepackage{amssymb}
\usepackage{amsmath}
\usepackage{lipsum}
\usepackage{csquotes}

\usepackage{fontenc}
\usepackage[utf8]{inputenc}
\usepackage{orcidlink}
\usepackage{ulem}
\usepackage{xcolor}
\usepackage{booktabs}

\journal{Physics Letters B}

\setcitestyle{numbers,square}

\begin{document}

\begin{frontmatter}




\title{Measuring $^{19,20}$O(p,n)$^{19,20}$F reactions using an active target detector}

\author[a]{Rohit Kumar\,\orcidlink{0000-0002-0450-7218}}
\author[a]{H. Desilets\,\orcidlink{0009-0009-4348-6526}}
\author[a]{R.~T. deSouza\,\orcidlink{0000-0001-5835-677X}}

\affiliation[a]{%
Department of Chemistry and Center for Exploration of Energy and Matter, Indiana University
2401 Milo B. Sampson Lane, Bloomington, Indiana 47408, USA}%

\begin{abstract}
Proton capture on $^{19,20}$O nuclei is measured in inverse kinematics with the active target detector MuSIC@Indiana using CH$_4$ as the target gas. Rejection of unreacted and inelastically scattered beam, along with transfer and fusion on the $^{12}$C allows extraction of the (p,n) cross section. As the cross-section for direct (p,n) processes at these energies is small, the measurement provides access to the proton fusion cross-section. An analysis approach that allows extraction of the proton fusion cross-section is detailed.

\end{abstract}



\begin{keyword}
Proton capture \sep Fusion \sep Radioactive beams \sep Active target detector \sep Multi-Sampling Ionization Chamber \sep Resonant behavior in fusion 



\end{keyword}

\end{frontmatter}




\section{Introduction}
\label{introduction}
Neutron and proton capture reactions have played a central role in nuclear physics since its inception and remain a topic of both fundamental and practical interest. Observation of neutron resonances following neutron capture on the heavy nuclei was instrumental in realizing that the decay of these nuclei could be treated in a semiclassical fashion \cite{Anderson50, Fenstermacher59, Kennett58}. Key to this treatment is the concept of the level density parameter \cite{Hurwitz51, Gadioli68, Greaves25}, intrinsic in a Hauser-Feshbach description of the statistical decay of an excited nucleus.  Proton capture on light nuclei is also interesting as in astrophysical environments these reactions are responsible for the synthesis of the lightest isotopes of an element \cite{Arnould03}. 
The advent of radioactive beam facilities and active target techniques opens a new era in the investigation of proton and neutron-capture reactions.
In the present analysis, using an active-target detector, we introduce an approach to measure (p,n) reactions enabling the study of proton capture. We specifically demonstrate this approach with the reaction $^{19,20}$O + p at 1 MeV$\leq$E$_{cm}$$\leq$2 MeV.

\begin{figure}
\begin{center}
\includegraphics[width=0.43\textwidth]{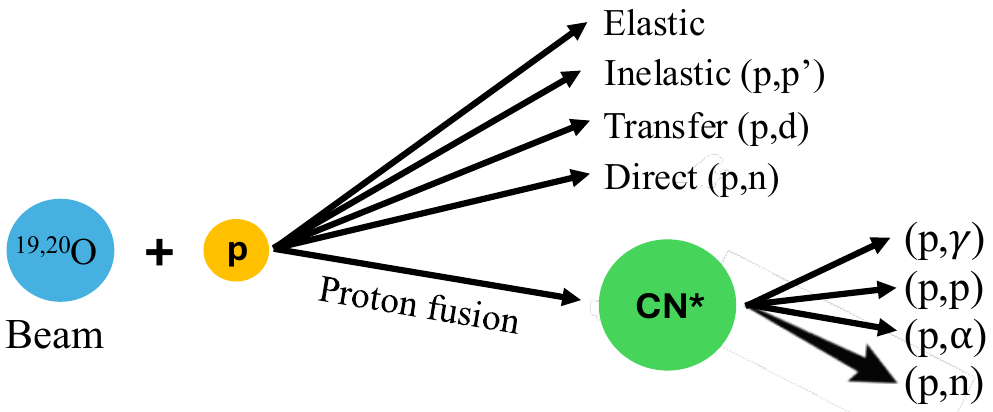}
\caption{Schematic illustration of the different reaction channels.}
\label{fig:Schematic}
\end{center}
\end{figure}

Different reaction channels that can occur when an oxygen nucleus and proton collide are schematically illustrated in Fig.~\ref{fig:Schematic}. The incident O ion can undergo elastic or inelastic scattering, transfer, a direct (p,n) reaction, or fusion leading to the formation of an excited compound nucleus (CN*). By identifying a F nucleus as the heavy reaction product, the present analysis provides a measurement of the (p,n) cross section. However, distinguishing the direct (p,n) process from neutron emission following fusion is challenging, requiring measurement of the neutron angular distribution to separate the peaked angular distribution associated with direct processes from the isotropic distribution indicative of compound nucleus formation.  Given the typically low intensity of the radioactive beams, high efficiency neutron detection is mandated for such a measurement -- a significant challenge. This challenge is presently mitigated as fusion dominates the direct (p,n) process at the energies investigated.
Calculations in TALYS for $^{20}$O + p
demonstrate that the cross section for fusion, $\sigma_{comp}$, is essentially
the same as the total reaction cross section, $\sigma_{rec}$, indicating the dominance of fusion at the energies measured
\cite{Koning23}.The deBroglie wavelength of a 1 MeV proton is $\sim$28 fm, larger than the nuclear diameter hence scattering of the incident proton on bound nucleons is minimal. Therefore the measured proton capture cross-section can be considered the proton fusion cross-section for the remainder of this manuscript.

Following fusion the excited compound nucleus can de-excite by $\gamma$, proton, $\alpha$, or neutron emission. For neutron-rich incident ions such as $^{20}$O, at the incident energies studied single neutron emission is heavily favored. Consequently, one effectively studies both proton capture on the ground-state $^{20}$O or, through microscopic reversibility, neutron-capture on the excited $^{20}$F providing important tests for nuclear reaction theories.

For the neutron-rich $^{19,20}$O in this work the single neutron emission channel dominates due to the low separation energies for a neutron as compared to an $\alpha$-particle, 6.6 and 8.1 MeV for $^{20}$F
and 8.1 and 10.3 MeV for $^{21}$F. The separation energies for a proton exceed those of an $\alpha$-particle, further suppressing proton emission.


Presented in Table\ref{table:decay_chan} are the de-excitation channels calculated following fusion for $^{19,20}$O in the Hauser-Feshbach statistical decay model code GEMINI++ \cite{Charity10}. In the case of $^{20}$O, single neutron emission corresponds to $>$99.8$\%$ of the cross-section while in the case of $^{19}$O some $\alpha$-emission reduces the single neutron emission to as low as 92$\%$ at the highest energies measured. Thus, for $^{20}$O, the measurement of the (p,n) cross section corresponds directly to the fusion cross section. The presence of other decay channels  for $^{19}$O means that extraction of the fusion cross section requires either accounting for the influence of these channels or their direct measurement. 

\begin{table} [t!]

  \begin{center}
    \begin{tabular}{cccccc}
    \hline
    {CN} & {E$^*$ (MeV)} & {n(\%)} & {$\alpha$(\%)} &  {$\alpha$n(\%)} & {p(\%)}\\
    \hline
    $^{20}$F & 12.1 & 98.2 & 1.8 & 0.0 & 0.0 \\
    $^{20}$F & 12.6 & 92.0 & 2.3 & 5.7 & 0.0 \\
    $^{21}$F & 12.2 & $>$99.9 & 0.0 & 0.0 & 0.0 \\
    $^{21}$F & 13.3 & 99.8 & 0.1 & 0.0 & 0.1 \\
    
   \bottomrule
    \end{tabular}
  \end{center}
  \caption{Yield of decay channels predicted by GEMINI++ \cite{Charity10} following proton fusion on $^{19,20}$O.}
  \label{table:decay_chan} 
\end{table}

\section{Experimental details}

Multi-Sampling Ionization Chamber (MuSIC) detectors are a powerful tool in investigating nuclear reactions with radioactive beams \cite{Johnstone21, Carnelli15, Blankstein23, Asher21b}. These detectors have been used to study ($\alpha$, p) \cite{Avila17, Jayatissa22} and ($\alpha$,n) \cite{Avila17, Blankstein24, Ong22} reactions of astrophysical interest as well as heavy-ion fusion \cite{Desilets25a, Carnelli14, Asher21a}. 

The radioactive beams in the experiment were provided by the SPIRAL1 facility of the GANIL accelerator complex in Caen, France.
The $^{20}$O (t$_\frac{1}{2}$ = 13.5 s) and $^{19}$O (t$_\frac{1}{2}$ = 26.5 s) beams were produced by bombarding a graphite target with a primary beam of $^{22}$Ne at E/A = 80 MeV. The resulting $^{20}$O (or $^{19}$O) ions were then accelerated by the CIME cyclotron to an energy of E/A = 2.7 MeV (3.0 MeV), selected in B$\rho$ by the ALPHA spectrometer, and transported to the experimental setup.

The principal detector used in the present work was the active-target detector, MuSIC@Indiana. 
Beam impinged on the detector at an intensity up to $\sim$1-2$\times$10$^4$ ions/s.
MuSIC@Indiana is a transverse-field ionization chamber designed for simultaneous measurement of beam and 
heavily-ionizing products within the detector gas volume. The anode is divided into twenty 12.5 mm-wide strips 
oriented perpendicular to the beam axis, which allows the ionization of each event to be tracked as the beam  
or heavily-ionizing reaction products traverse the gas. 
A distinguishing feature of MuSIC@Indiana is the ability to insert a silicon detector from downstream directly into the active volume. This capability allows the direct energy loss measurement for ions traversing the gas.
In the present experiment, the detector was filled with high-purity CH$_{4}$ at pressures between 110 and 130 torr. Detailed descriptions of the detector design and performance can be found in Refs.~\cite{Johnstone21, Desilets25}.

 \begin{figure}[h]
 \begin{center}
\includegraphics[width=0.45\textwidth]{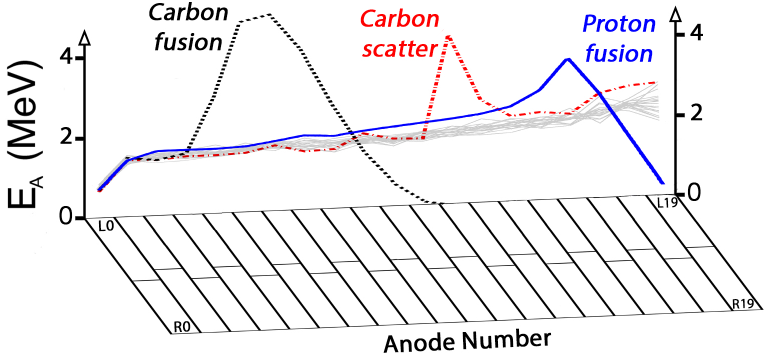}
\caption{Typical traces for fusion on carbon (black, dotted), two-body scattering on carbon (red, dash-dot), and proton fusion (blue, solid) events are superimposed on the anode structure of MuSIC@Indiana. Multiple beam traces (light gray) are shown for reference.}
\label{fig:Traces}
\end{center}
\end{figure}

As an incident ion traverses the detector, the energy deposited on the anode strips is recorded. The set of measured anode-strip energies, E$_{\rm A}$, is collectively referred to as a \enquote{trace}.
Representative traces, indicative of the ionization produced by the beam, as well as three distinct processes, are presented in 
Fig.~\ref{fig:Traces} for the $^{20}$O beam.
The light-gray traces indicate ionization due to the passage of beam ions.
In this case, the energy deposited on an anode increases from $\sim$1.3 MeV on entering the detector to $\sim$2 MeV at the detector exit. 

The most distinctive trace evident in Fig.~\ref{fig:Traces}, is indicated by a dotted (black) trace. It corresponds to a dramatic increase in deposited energy occurring at anode 4.
For several anodes after the initial increase in ionization, a significantly larger ionization as compared to the beam is observed. This increased ionization indicates the presence of  
a heavily ionizing fusion product formed from the fusion of the incident O ion with a $^{12}$C nucleus. This evaporation residue (ER) exhibits increased ionization upto the Bragg maximum followed by a general monotonic decrease.
The increased ionization at anode 4 reveals the position and thus energy at which the fusion occurred.
The fusion excitation functions for $^{19,20}$O + $^{12}$C have been previously reported \cite{Desilets25a,Desilets25}.

Indicated in Fig.~\ref{fig:Traces} are two other reactions that occur.
The dash-dot (red) trace is representative of two-body scattering events (e.g., inelastic scattering, transfer, etc.) with a $^{12}$C nucleus.
Although these traces initially exhibit a spike in ionization, they are clearly distinguished from fusion by the return to a beam-like ionization after the initial peak. Replacing the detector gas with hydrogen would eliminate both the two-body scattering from carbon as well as carbon fusion providing a cleaner measurement for proton fusion.
Two-body scattering from a proton is similar to that of carbon although the energy deposit is less in magnitude. 

Also observed in Fig.~\ref{fig:Traces} is a solid (blue) trace that exhibits only a modest increase in ionization relative to the beam. This increased ionization is observed over several anodes and is consistent with only a small change in atomic number, e.g. proton fusion.
Fusion of a proton with an $^{20}$O ion results in formation of $^{21}$F$^*$ with an excitation energy 12.2 MeV $\leq$ E$^*$ $\leq$ 13.3 MeV.
Due to the lower neutron binding energy, 8.1 MeV, as compared to that of both a proton or an $\alpha$-particle, 11.1 MeV and 10.3 MeV respectively, the $^{21}$F preferentially decays to $^{20}$F. By distinguishing F from O ions, measurement of the (p,n) cross section is accomplished. 
As the $^{20}$F travels through the detector, the difference in ionization between it and the un-reacted beam increases. 
This characteristic behavior of the F traces enables (p,n) events to be distinguished from events associated with un-reacted beam, two-body scattering from $^{12}$C or $^{1}$H, and fusion with $^{12}$C.

\begin{figure}[h]
\begin{center}
\includegraphics[width=0.45\textwidth]{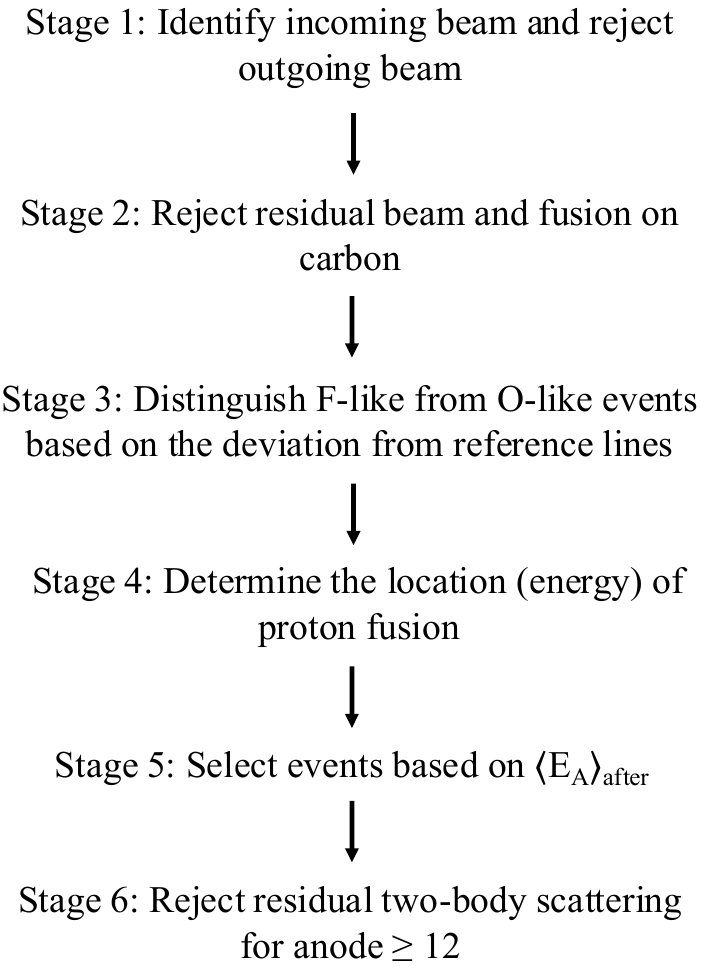}
\caption{Flowchart depicting the analysis logic utilized.}
\label{fig:Flowchart}
\end{center}
\end{figure}

\section{Analysis framework} 
The challenge faced in isolating proton fusion with radioactive beams using an active target detector, namely the minimal separation of the reaction product with $\Delta$Z = 1 from beam, is similar to the problem confronted in investigating ($\alpha$,p) reactions \cite{Jayatissa22} and somewhat more challenging than the study of ($\alpha$,n) reactions \cite{Avila17, Blankstein24} which have $\Delta$Z = 2.

A flowchart of the analysis used to determine the (p,n) cross-section for $^{19,20}$O + p reactions is shown in Fig.~\ref{fig:Flowchart}. After identification of the incident beam and the rejection of outgoing beam, events corresponding to fusion on carbon are rejected based on the maximum ionization observed. Comparison of a measured trace to the average measured beam trace and a predicted F reference trace provides further selection.
Events corresponding to (p,n) reactions are subsequently selected by their consistent increased ionization relative to the beam. Each of these steps is described in detail below.

\subsection{Stage 1: Identify incoming beam and reject outgoing beam}

The first step of the analysis, as indicated in Fig.~\ref{fig:Flowchart}, is the 
identification of each incident ion as the desired beam of interest, either $^{19}$O or $^{20}$O. Identification was accomplished by utilizing the first four anode strips of MuSIC@Indiana for a E$_{A0}$ vs E$_{(A1+A2+A3)}$ measurement. This selection not only eliminates possible contaminants in the beam but also rejects any reactions originating from the mylar window which separates the gas in MuSIC@Indiana from the high-vacuum section upstream. 

\begin{figure}[h]
\begin{center}
\includegraphics[width=0.35\textwidth]{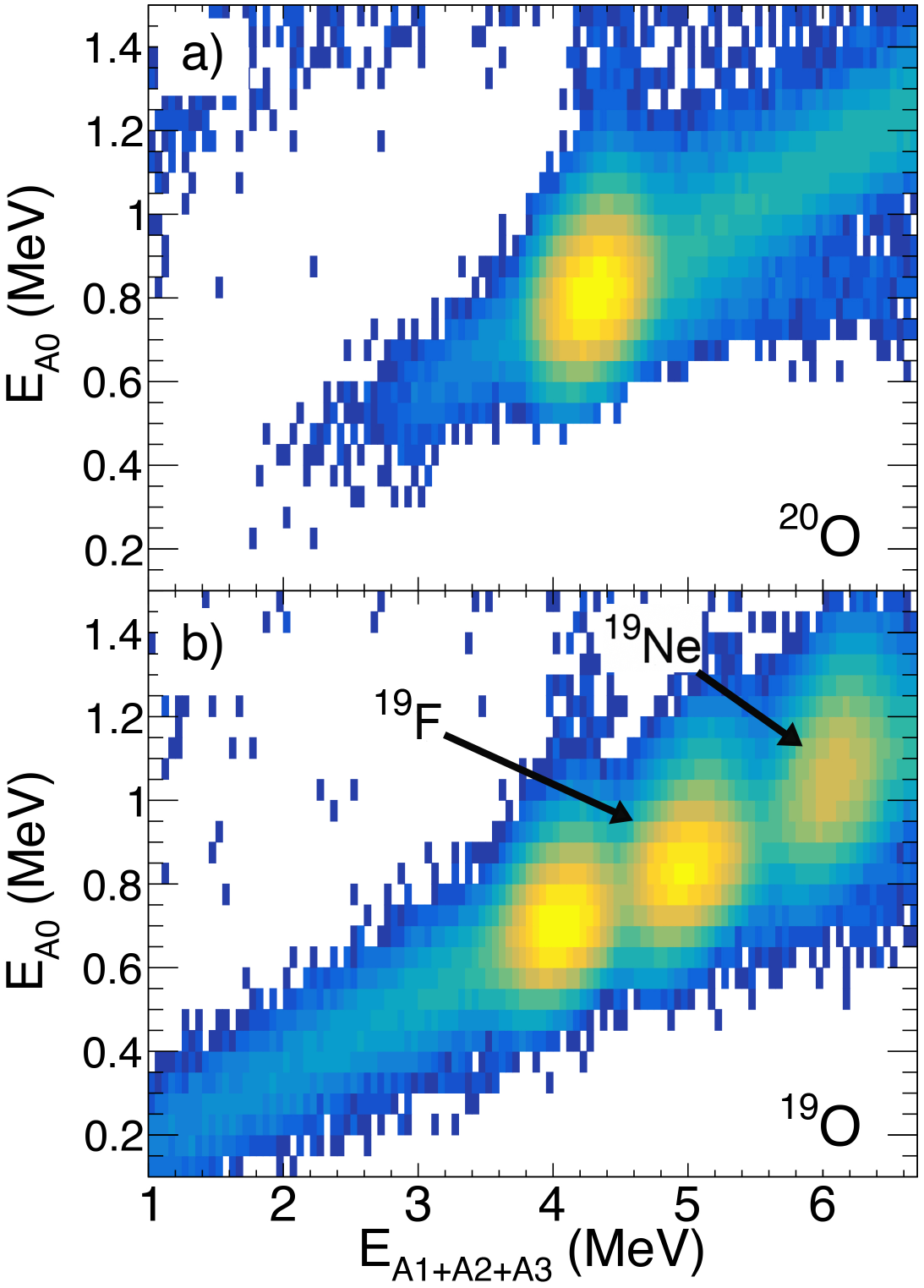}
\caption{Identification of the incident beam based upon the energy deposit of the first four anodes of MuSIC@Indiana. Panel a) For incident $^{20}$O beam, no evidence of beam contamination is observed. Panel b) Three peaks are observed corresponding to the presence of $^{19}$F ($\sim$37\%) and $^{19}$Ne ($\sim$4\%) along with $^{19}$O beam. However, when averaging over all the measured dataset $^{19}$F is $\sim$55\% and $^{19}$Ne remains unchanged.}
\label{fig:PID}
\end{center}
\end{figure}

Presented in Fig.~\ref{fig:PID} is the identification of the incident beam based upon the energy deposit of the first four anodes of MuSIC@Indiana. In Fig.~\ref{fig:PID}a, for $^{20}$O, a single dominant peak is observed. In addition to the peak a locus is evident extending to higher energy deposit. This locus corresponds to pileup events in the detector which occur at the 1$\%$ level.  On the low energy deposit side of the main peak a slight tail in the distribution is also noted.

In Fig.~\ref{fig:PID}b in contrast to the case of $^{20}$O,
three peaks are observed for $^{19}$O. These peaks indicate the presence of $^{19}$F and $^{19}$Ne contaminants at a significant level along with the $^{19}$O beam. The identity of these peaks is based on the measured energy loss as well as the B$\rho$ of the beamline. The  $^{19}$F and $^{19}$Ne are present at a level of $\sim$37\% and $\sim$4\% respectively. Both contaminants are well separated from the $^{19}$O beam of interest. It should be noted that the low energy deposit tail for $^{19}$O ($<$ 0.2\%) is more prominent as compared to $^{20}$O. We attribute these low energy deposit events with ionization due to incident light charged particles produced upstream by a poorer beam tune. As the identity of the incoming ion is selected by gating on the 
$^{19}$O peak, neither the contaminant beams nor the low energy tail pose an issue for the subsequent analysis.

The left/right segmentation of the first four anodes was used to ensure that the incident beam was centered at the entrance of the detector. Outgoing beam was rejected by examining the characteristic energy loss measured in E$_{A18}$ vs E$_{A19}$. This rejection  thus selected
events in which an incident ion had either reacted or scattered in the detector. 

\subsection{Stage 2: Reject residual beam and fusion on carbon}

In order to focus on proton fusion, events corresponding to beam are rejected by requiring E$_{A}^{max}$ $>$ E$_{th,min}$, where
E$_{A}^{max}$ is the maximum anode energy deposited within a trace. Fusion of the incident beam with carbon is rejected by requiring that E$_{A}^{max}$ $<$ E$_{th,max}$. For $^{20}$O at P=130 torr, E$_{th,min}$ = 2 MeV and E$_{th,max}$ = 3.5 MeV.

\subsection{Stage 3: Distinguish F-like from O-like events based on the deviation from reference lines}

\begin{figure}[h]
\begin{center}
\includegraphics[width=0.45\textwidth]{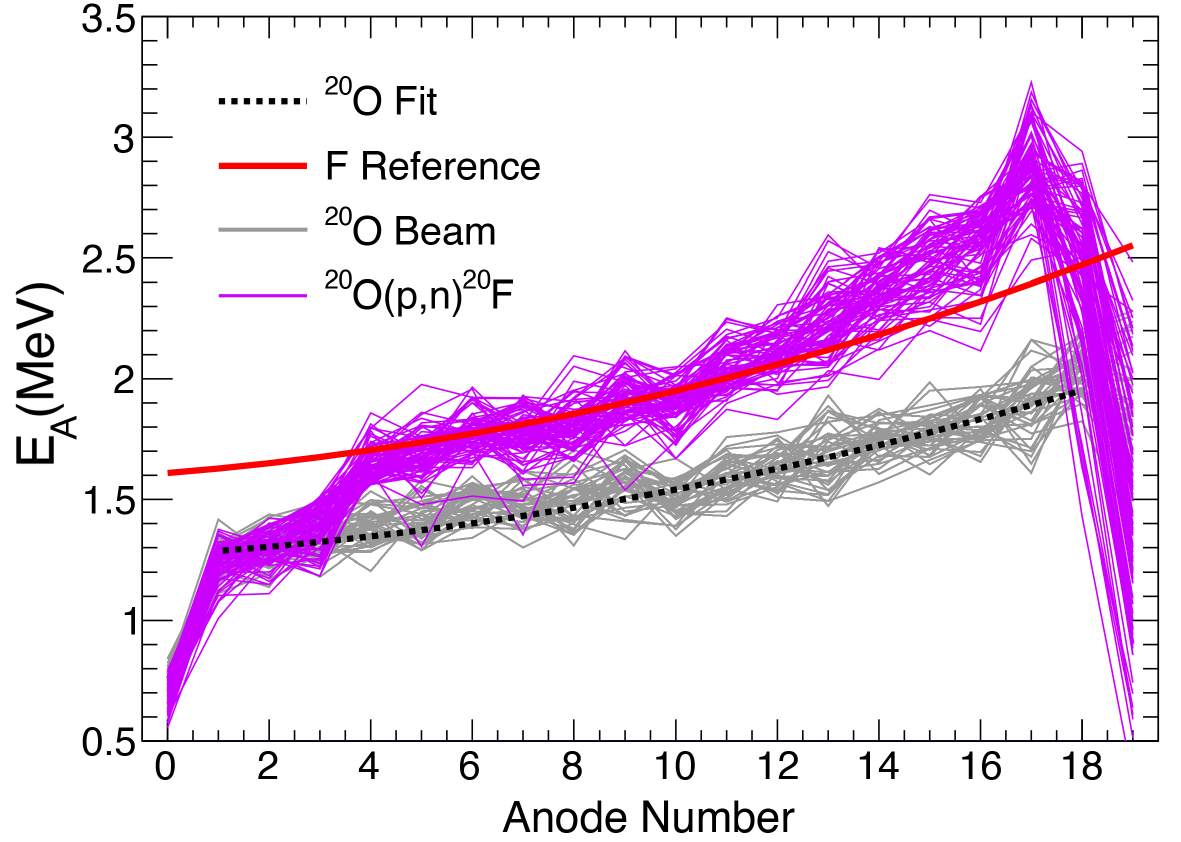}
\caption{Putative proton fusion trace (solid, magenta) is compared to a few beam traces (solid, gray). The average of the beam traces used as the dotted (black) line indicates the reference beam trace. Indicated by the solid (red) line, based upon a 
Z$^2$ scaling is the reference line for F ions.}
\label{fig:OFAvg}
\end{center}
\end{figure}

Events are subsequently designated as more \enquote{O-like} or \enquote{F-like}.
This classification was realized by calculating the deviation of a given trace from O and F reference traces. The reference trace for O was calculated by fitting the traces of representative beam events, defined by the  E$_{A18}$ vs E$_{A19}$ selection, with a second-order polynomial. 
Shown in Fig.~\ref{fig:OFAvg} are some representative beam traces along with reference line for O. 
To calculate the reference line for F, the Z$^2$ dependence of the energy deposit for ions of the same velocity was assumed, consistent with the Bethe-Bloch formula \cite{Bethe30, Bloch33}.
The reference F line along with candidate proton fusion traces which manifest increased ionization after anode 4 and have maximum ionization in anode 17 are also presented in Fig.~\ref{fig:OFAvg}.

The reasonably good agreement of the traces to the F reference for several anodes after its initial increase from beam ionization suggests that the these traces corresponds to F ions.  For these candidate proton fusion events, the resulting ER reaches a maximum energy deposit in anode 17 before stopping in the detector volume. It can be observed that, as expected, close to the Bragg maximum, the Z$^2$ scaling is a poor description.

For a given trace, the deviation from the F reference line, $D_F$, was calculated as:
\begin{equation}
D_F=\sum_{i=A_{th}}^{A_{max}}{(E_A(i)-E_{A, Fref}(i))^2}
\end{equation}

where $A_{th}$ is the first anode whose energy deposit is 150 KeV above the beam average. The summation is performed up to the anode which contains the maximum energy deposit of the trace, $A_{max}$. The deviation from the O reference, D$_O$, was calculated analogously. 

\begin{figure}[h]
\begin{center}
\includegraphics[width=0.45\textwidth]{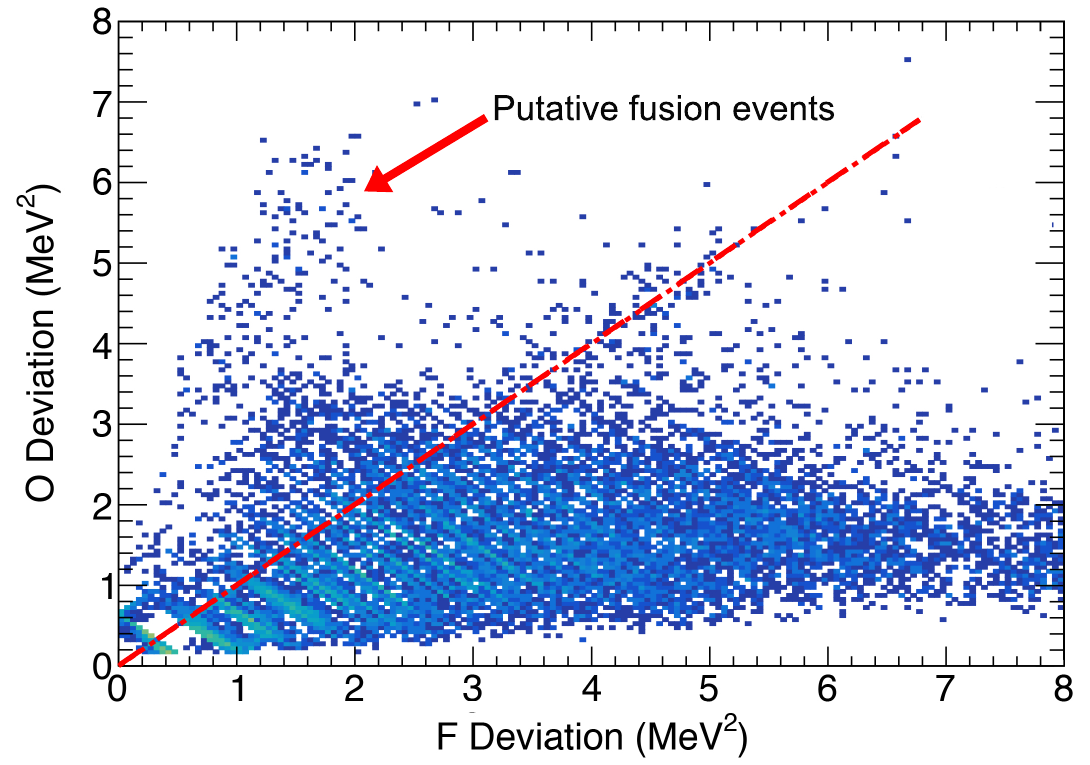}
\caption{Deviation of each event trace from the average O and F reference lines. Events lying to the right of the diagonal line $\sim$90\% are rejected. The putative fusion events cluster at low F deviation and high O deviation.}
\label{fig:OFDev}
\end{center}
\end{figure}

Fig.~\ref{fig:OFDev} shows the deviation of each event at this stage from O and F reference lines. The structure of the data into a set of anti-correlated loci is due to the finite anode structure of the detector. Each anti-correlated locus corresponds to the initial deviation from beam occurring at a specific anode. Most of the events exhibit a smaller deviation from O than F. Well separated from most of the events is a small group that manifests a smaller deviation from F than from O, hence are putative proton fusion events. The diagonal dashed (red) line in Fig. ~\ref{fig:OFDev} separates events into those that are more \enquote{O-like} or \enquote{F-like}. By selecting traces that only lie above the dashed (red) line $\sim$ 90\% of beam-like events are rejected. From Fig.~\ref{fig:OFDev} it is clear that some beam-like events which lie close to the diagonal line are retained. These events are effectively rejected in subsequent stages.

\subsection{Stage 4: Determine the location (energy) of proton fusion}

The location at which the proton fusion occurs was determined by comparing E$_A$ with an anode dependent threshold, E$_{th}$. These thresholds were determined by fitting the beam E$_A$ distribution for each anode with a Gaussian. The Gaussian centroids were then fit with a second-order polynomial depicted as the average beam line in Fig.~\ref{fig:OFAvg}. The threshold, E$_{th}$, was then determined by adding 190 keV ($\sim$3$\sigma$) to the beam average in order to account for the energy dispersion (FWHM) of the beam. 

\begin{figure}[h]
\begin{center}
\includegraphics[width=0.35\textwidth]{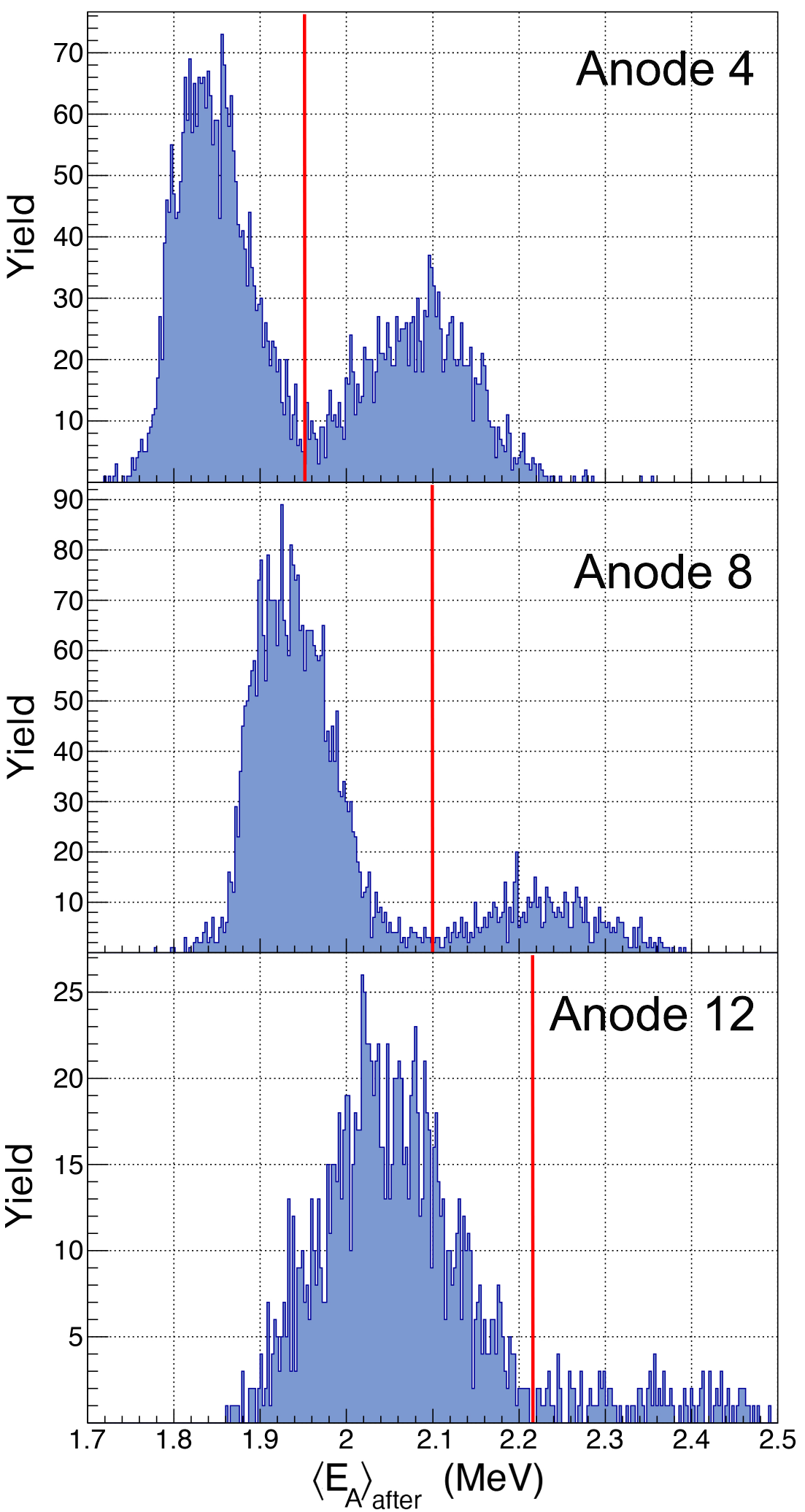}
\caption{Distribution of $<$E$>$ $_{after}$. Two peaks are visible: the left peak corresponds to scattered beam events and the right peak corresponds to proton fusion events. The beam scattered events are rejects by the vertical (red) line.}
\label{fig:Eafter}
\end{center}
\end{figure}

\subsection{Stage 5: Select events based on $<$E$>$$_{after}$}

To further select proton fusion, specifically isolating it from proton two-body scattering (elastic or inelastic), the quantity $\langle$E$\rangle$$_{after}$ was constructed.
The quantity $\langle$E$\rangle$$_{after}$ is the average energy deposit from the anode where E$_{th}$ is exceeded up to the anode where E$_{max}$ occurs. The distribution of $\langle$E$\rangle$$_{after}$ for anodes 4, 8 and 12 is presented in Fig.~\ref{fig:Eafter}.
In all cases, a two-peaked distribution is evident. To determine the nature of the events associated with each peak, individual traces were examined after selecting on either the lower or higher peak in $\langle$E$\rangle$$_{after}$. Traces associated with the lower peak manifested the characteristic energy deposit of two-body scattering from a proton while traces associated with the larger 
 $\langle$E$\rangle$$_{after}$
were associated with proton fusion. The solid (red) line indicates the minimum value used to select proton fusion events.
The bimodal nature of the  $\langle$E$\rangle$$_{after}$ distribution is clearly evident for anodes prior to 12 as evident in Fig.~\ref{fig:Eafter} allowing selection of proton fusion for these anodes. 
For anode 12 and 13 distinguishing fusion from two-body scattering becomes difficult as the cross section for two-body grows and the fusion cross section decreases. To isolate the fusion cross section for these last two anodes required additional analysis. The loss of separation between these two reactions in $\langle$E$\rangle$$_{after}$ beyond anode 13 for $^{20}$O and anode 10 in $^{19}$O provided the lower limit over which the cross section was measured.

\subsection{Stage 6: Reject residual two-body scattering for anode $\ge$ 12}

\begin{figure}[h]
\begin{center}
\includegraphics[width=0.45\textwidth]{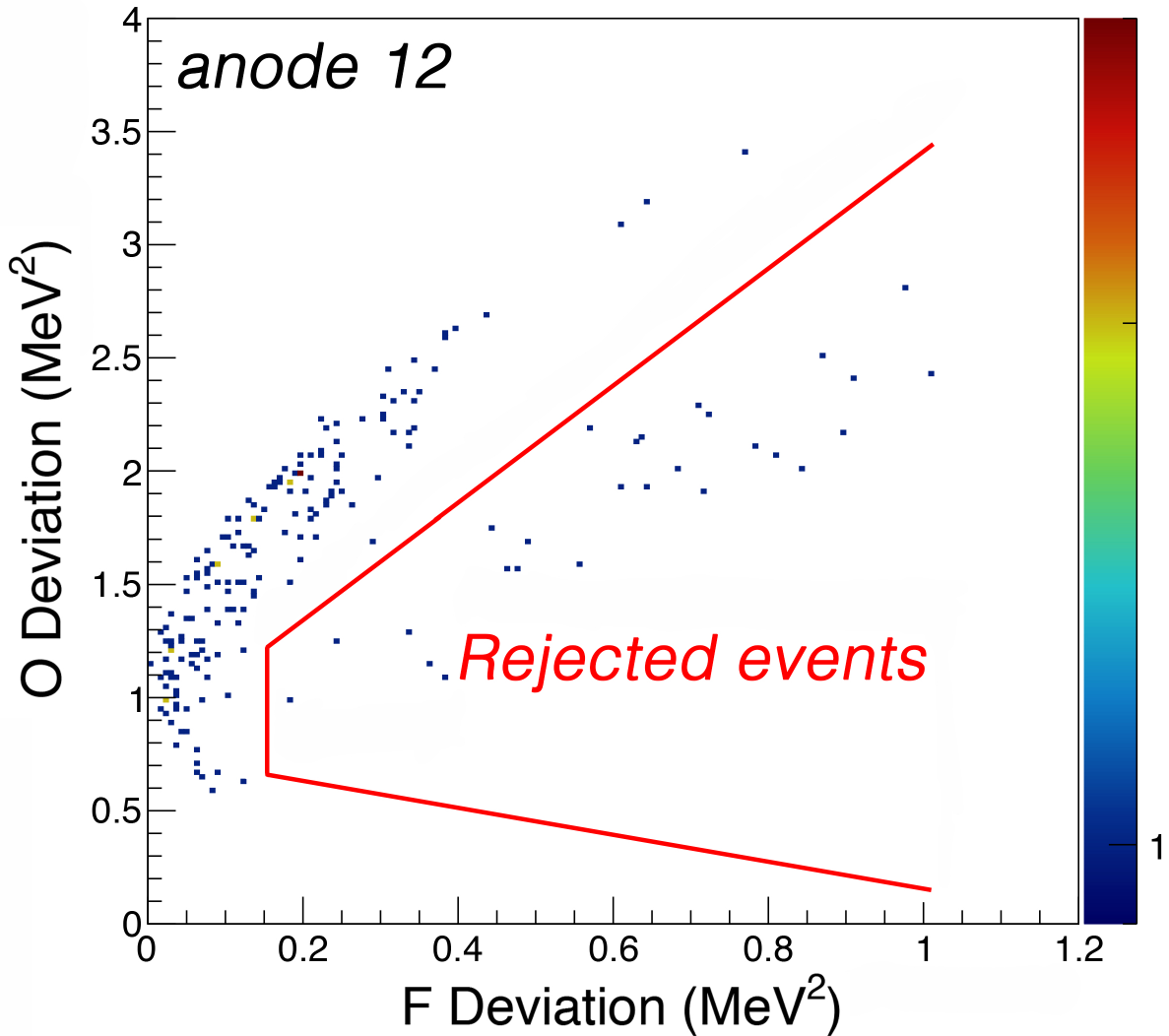}
\caption{Correlation between the O-deviation and the  F-deviation for events past stage 5. Events that occur right to the parabola are rejected. This stage of the analysis was only used for anode $\geq$ 12.}
\label{fig:Fd_Od}
\end{center}
\end{figure}

A final selection of proton fusion events is realized by re-examining the deviation from the F and O reference lines. In contrast to the prior use of the deviation, here the deviation is only calculated from the first anode past threshold for five subsequent anodes. Restricting the number of anodes in the calculation, mitigates the impact of the deviation of the Bragg peak from the reference lines. In Fig.~\ref{fig:Fd_Od}, the correlation between the two deviations is examined. Most of the data falls on a parabolic locus. The minimum of the parabola is located at D$_F$$\sim$0 and D$_O$$\sim$1 MeV$^2$. 
The offset in D$_O$ corresponds to the separation of the O and F lines summed over five anodes. Events which lie well off the locus are rejected. We emphasize that this final stage of rejection is only necessary deep in the detector where the relative importance of two-body scattering increases.

\section{Calculation of the proton fusion cross section}

After isolating proton fusion events, N$_{ER}$, from beam and other reaction types, the cross-section can be determined. The fusion cross-section, $\sigma_F$ is related to the number of proton fusion events, the number of incident beam ions, N$_{Beam}$ and the thickness of an anode strip, $\Delta$x by: $\sigma_F = N_{ER}/(N_{Beam}\times\Delta x)$. 
Complete detection of the ER within the active volume of MuSIC@Indiana eliminates the need for an efficiency correction. Unlike thin-target measurements, which are sensitive to the measured ER angular distribution,  MuSIC measurements intrinsically provide an angle-integrated measurement of the cross-section. The uncertainty in the measured cross-section is largely determined by the statistical uncertainty associated with $N_{ER}$. 
In addition, a systematic uncertainty of 5$\%$ associated with the mis-identification of fusion events as non-fusion events for $N_{ER}$ is included.  The uncertainty associated with $N_{Beam}$ is negligible, and the uncertainty associated with $\Delta x$ is defined by the pressure variation in MuSIC@Indiana. All these uncertainties are included in the reported error bars.
The finite size of each anode strip results in an uncertainty in the energy at which the reaction occurs.

\begin{figure}
\begin{center}
\includegraphics[width=0.45\textwidth]{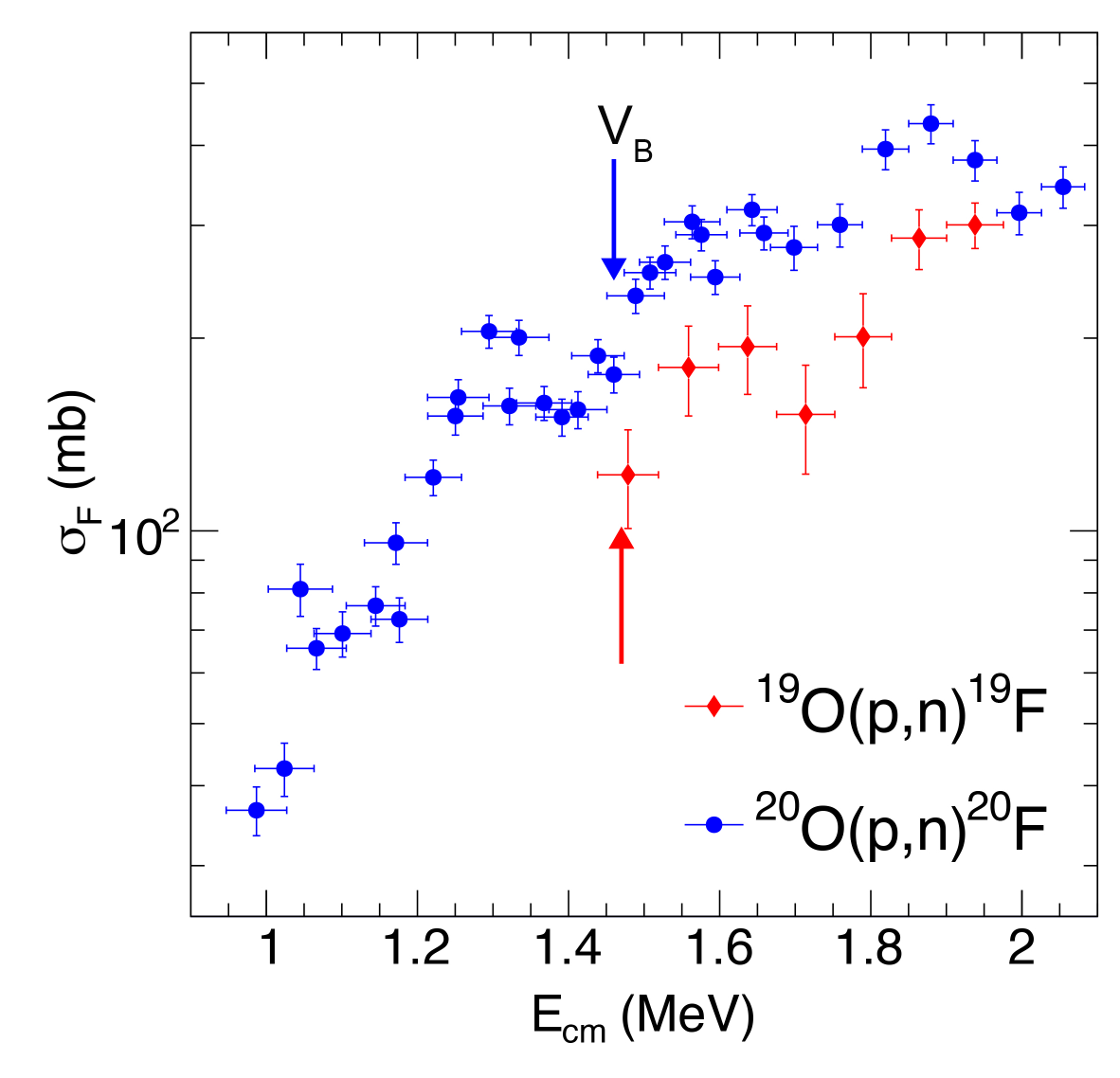}
\caption{Dependence of the cross section for $^{19}$O(p,n) and $^{20}$O(p,n) on E$_{cm}$. The Bass barrier, V$_{B}$ \cite{Bass74} for each system is indicated by the arrows: V$_{B}$ = 1.47 MeV for $^{19}$O and V$_{B}$ = 1.46 MeV for $^{20}$O.}
\label{fig:XsectErr}
\end{center}
\end{figure}

In Fig.~\ref{fig:XsectErr} the measured excitation functions for $^{20}$O(p,n) and $^{19}$O(p,n) are presented. The excitation function for $^{20}$O is measured over the interval 1.0 $\leq$ E$_{cm}$ $\leq$2.05 MeV reaching a cross section level of $\sim$40 mb at the lowest energy. In the case of $^{19}$O a smaller interval, 1.48 $\leq$ E$_{cm}$ $\leq$1.94 MeV, was measured. 
Consistent with a barrier-governed process, one observes a general decrease in the cross-section, $\sigma_F$.

For $^{20}$O a non-smooth behavior is observed with an indication of three broad peaks superimposed on the excitation function. In the case of $^{19}$O one also observes a non-smooth behavior, although the statistical quality of the data is somewhat worse. The slight suppression in $\sigma_F$ for $^{20}$O at E$_{cm}$$\sim$1.7 MeV is also reflected in a suppression of $\sigma_F$ for $^{19}$O at approximately the same energy. These broad peaks for $^{20}$O with a width of 50 - 100 keV have been associated with the presence of short-lived quasi-bound states in the compound nucleus \cite{Desilets25b}.

\section{Conclusions}
Use of the active target detector MuSIC@Indiana with CH$_{4}$ gas and incident radioactive beams of $^{20}$O and $^{19}$O ions enabled efficient extraction of the (p,n) cross section. 
Separation of F reaction products from the unreacted O ions made isolation of the (p,n) cross-section possible. This separation of $\Delta$Z = 1 is also relevant for study of ($\alpha$,p) and ($\alpha$,n) reactions.

The two contributions to the cross section are direct (p,n) reactions and proton fusion. Although at the measured incident energies, proton fusion is expected to dominate over direct (p,n) reactions \cite{Koning23}, determining the extent to which direct processes influence the measured cross section will require further experiments capable of measuring the neutron energy and angle.

Measurement of this proton fusion cross-section, $\sigma_F$, is facilitated by the dominant decay of the compound nucleus via emission of a single neutron. The analysis technique presented allows proton fusion to be effectively distinguished from both unreacted beam and other competing reactions. 
Accurate determination of the proton fusion location results in measurement of the excitation function. The extracted excitation function for $^{20}$O manifests a clear oscillatory behavior with broad peaks indicative of short-lived states. A non-smooth excitation function is also measured for proton fusion on $^{19}$O. These initial measurements demonstrate an effective means to investigate proton fusion with neutron-rich radioactive beams.

\section{Acknowledgments}
We acknowledge the high-quality beam and experimental support provided by the technical and scientific staff at the Grand Acc\'{e}l\'{e}rateur National d'Ions Lourds (GANIL), in particular D. Allal, B. Jacquot, and D. Gruyer. The contributions of our collaborators, S. Hudan, C. Ciampi, A. Chbihi, K.~W. Brown in acquiring the data is gratefully acknowledged.
We are thankful for the high-quality services of the Mechanical Instrument Services and Electronic Instrument Services facilities at Indiana University.

This work was supported by the U.S. Department of Energy Office of Science under Grant No. 
DE-SC0025230 and Indiana University. This research has received funding from the European Union's HORIZON EUROPE Program under grant agreement no. 101057511. R. deSouza gratefully acknowledges the support of the GANIL Visiting Scientist Program.

 \textit{Data Availability Statements}. 
The supporting data for this article are from the e831\_{21} experiment and are registered as
https://doi.org/10.26143/GANIL-2023-E831\_21 following the GANIL Data Policy.




\bibliographystyle{elsarticle-num} 






\end{document}